\documentclass[aps,prb,twocolumn,showpacs,superscriptaddress]{revtex4}
\usepackage{graphicx}

\begin{document}

\title{Nature of the antiferromagnetic and nematic transitions in Sr$_{1-x}$Ba$_x$Fe$_{1.97}$Ni$_{0.03}$As$_2$}
\author{Dongliang Gong}
\affiliation{Beijing National Laboratory for Condensed Matter Physics, Institute of Physics, Chinese Academy of Sciences, Beijing 100190, China}
\affiliation{School of Physical Sciences, University of Chinese Academy of Sciences, Beijing 100190, China}
\author{Zhaoyu Liu}
\affiliation{Beijing National Laboratory for Condensed Matter Physics, Institute of Physics, Chinese Academy of Sciences, Beijing 100190, China}
\affiliation{School of Physical Sciences, University of Chinese Academy of Sciences, Beijing 100190, China}
\author{Yanhong Gu}
\affiliation{Beijing National Laboratory for Condensed Matter Physics, Institute of Physics, Chinese Academy of Sciences, Beijing 100190, China}
\affiliation{School of Physical Sciences, University of Chinese Academy of Sciences, Beijing 100190, China}
\author{Tao Xie}
\affiliation{Beijing National Laboratory for Condensed Matter Physics, Institute of Physics, Chinese Academy of Sciences, Beijing 100190, China}
\affiliation{School of Physical Sciences, University of Chinese Academy of Sciences, Beijing 100190, China}
\author{Xiaoyan Ma}
\affiliation{Beijing National Laboratory for Condensed Matter Physics, Institute of Physics, Chinese Academy of Sciences, Beijing 100190, China}
\affiliation{School of Physical Sciences, University of Chinese Academy of Sciences, Beijing 100190, China}
\author{Huiqian Luo}
\affiliation{Beijing National Laboratory for Condensed Matter Physics, Institute of Physics, Chinese Academy of Sciences, Beijing 100190, China}
\author{Yi-feng Yang}
\affiliation{Beijing National Laboratory for Condensed Matter Physics, Institute of Physics, Chinese Academy of Sciences, Beijing 100190, China}
\affiliation{School of Physical Sciences, University of Chinese Academy of Sciences, Beijing 100190, China}
\affiliation{Collaborative Innovation Center of Quantum Matter, Beijing 100190, China}
\author{Shiliang Li}
\email{slli@iphy.ac.cn}
\affiliation{Beijing National Laboratory for Condensed Matter Physics, Institute of Physics, Chinese Academy of Sciences, Beijing 100190, China}
\affiliation{School of Physical Sciences, University of Chinese Academy of Sciences, Beijing 100190, China}
\affiliation{Collaborative Innovation Center of Quantum Matter, Beijing 100190, China}

\begin{abstract}

We have systematically studied the antiferromagnetic and nematic transitions in Sr$_{1-x}$Ba$_x$Fe$_{1.97}$Ni$_{0.03}$As$_2$ by magnetic susceptibility and uniaxial-pressure resistivity measurements, respectively. The derivatives of the temperature dependence of both magnetic and nematic susceptibilities show clearly sharp peaks when the transitions are first order. Accordingly, we show that while both of the magnetic and nematic transitions change from first order to second order with increasing Barium doping level, there is a narrow doping range where the former becomes second order but the latter remains first order, which has never been realized before in other systems. Moreover, the antiferromagnetic and nematic transition temperatures become different and the jump of nematic susceptibility becomes small in this intermediate doping range. Our results provide key information on the interplay between magnetic and nematic transitions. Concerning the current debate on the microscopic models for nematicity in iron-based superconductors, these observations agree with the magnetic scenario for an itinerant fermionic model.

\end{abstract}


\maketitle

\section{INTRODUCTION}
The nature of the antiferromagnetic (AF) and nematic transitions in iron-based superconductors has attracted much interest. The underlying mechanism of both orders may be crucial to our understanding of superconductivity in these materials \cite{FernandesRM14}. The establishment of the electronic nematic order, which breaks the four-fold rotational symmetry of the underlying lattice, is always accompanied by a structural phase transition due to the symmetry constraint. Theoretical understanding of these phase transitions can be mainly divided into two groups based on the spin or orbital degree of freedom depending on the microscopic driving force of the nematic order. In the orbital scenario, the orbital ordering gives rise to the structural transition and then triggers the magnetic transition at the same or lower temperature \cite{LvW09,LeeCC09,BrydonPMR11,LeeWC13}. In the spin scenario, on the other hand, it has been argued that magnetic fluctuations are of primary responsibility for triggering the nematic instability, although it is still not clear whether a correct microscopic theory should be built solely on a local spin model or the itinerant characteristic of the Fe $3d$ electrons should be taken into full account \cite{KimMG11,FangC08,XuC08,QiY09,FernandesRM11,FernandesRM12,WysockiAL11,KamiyaY11,ApplegateR12}.

One way to test these theories is to carefully compare their predictions with experimental results. Especially, the nature of these transitions can reveal crucial information on the origin of the nematic order. The rich phase diagrams of the electron-doped ``122" systems, i.e., $A$Fe$_2$As$_2$ ($A$ = Ca, Sr, Ba) and its electron doped materials, give us an opportunity to do so. Both the magnetic and structural transitions in CaFe$_2$As$_2$ and SrFe$_2$As$_2$ are strongly first order and happen at the same temperature \cite{TorikachviliMS08,NiN08,ColombierE09}. For BaFe$_2$As$_2$, while the nature of these two transitions were initially under debate \cite{RotterM08,KitagawaK08,WilsonSD09,MatanK09}, further detailed studies have suggested that the structural transition is second order followed by a first order magnetic transition \cite{KimMG11,RotunduCR10}. Doping electron carriers (Co or Ni) into BaFe$_2$As$_2$ makes both transitions second order \cite{PrattDK09,NandiS10,HarrigerLW09,LuoH12,LuX13}. Therefore, a magnetic tricritical point has been suggested where the AF transition changes from first to second order with the structural transition remaining second order \cite{KimMG11,RotunduCR11}, but a nematic tricritical point has not been reported. Moreover, an intermediate phase where a first order nematic transition followed by a second order AF transition has never been found.

In this paper, we give detailed studies on the phase diagram of Sr$_{1-x}$Ba$_{x}$Ni$_{0.03}$Fe$_{1.97}$As$_2$ by measuring the resistivity, magnetic susceptibility and nematic susceptibility. It has been shown that Ba doping into SrFe$_2$As$_2$ can continuously reduce both the magnetic and structural transition temperatures \cite{MitchellJE12}. We doped 1.5\% of Ni into the system, where the AF and nematic transitions in the $x =$ 0 sample are still first order but those in the $x =$ 1 sample are clearly second order \cite{LuX16}. The crossover from first order to second order transition happens at different doping levels for magnetic and nematic systems. In this intermediate region, although the nematic transition is first order, the jump of the nematic susceptibility becomes significantly small. These observations are consistent with the theory based on the magnetic scenario for an itinerant fermionic model in excellent details \cite{FernandesRM12}.

\section{Experiments}
Single crystals of Sr$_{1-x}$Ba$_{x}$Ni$_{0.03}$Fe$_{1.97}$As$_2$ samples were grown by the self-flux methods as reported previously \cite{ChenY11}. The magnetic susceptibility measurements were carried out on the magnetic property measurement system (MPMS) at 7 Tesla applied within the a-b plane. Both the resistivity and nematic susceptibility were measured on the physical property measurement system (PPMS). The samples were cut into thin rectangular plates along the tetragonal (1,1,0) direction. After measuring the resistivity at zero pressure, the samples were glued on a home-made uniaxial pressure device as reported elsewhere \cite{LiuZ16}. The pressure is estimated from previous measurements \cite{LiuZ16}. In all measurements, the temperature was stabilized long enough to obtain reliable data.

\begin{figure}
\includegraphics[scale=0.225]{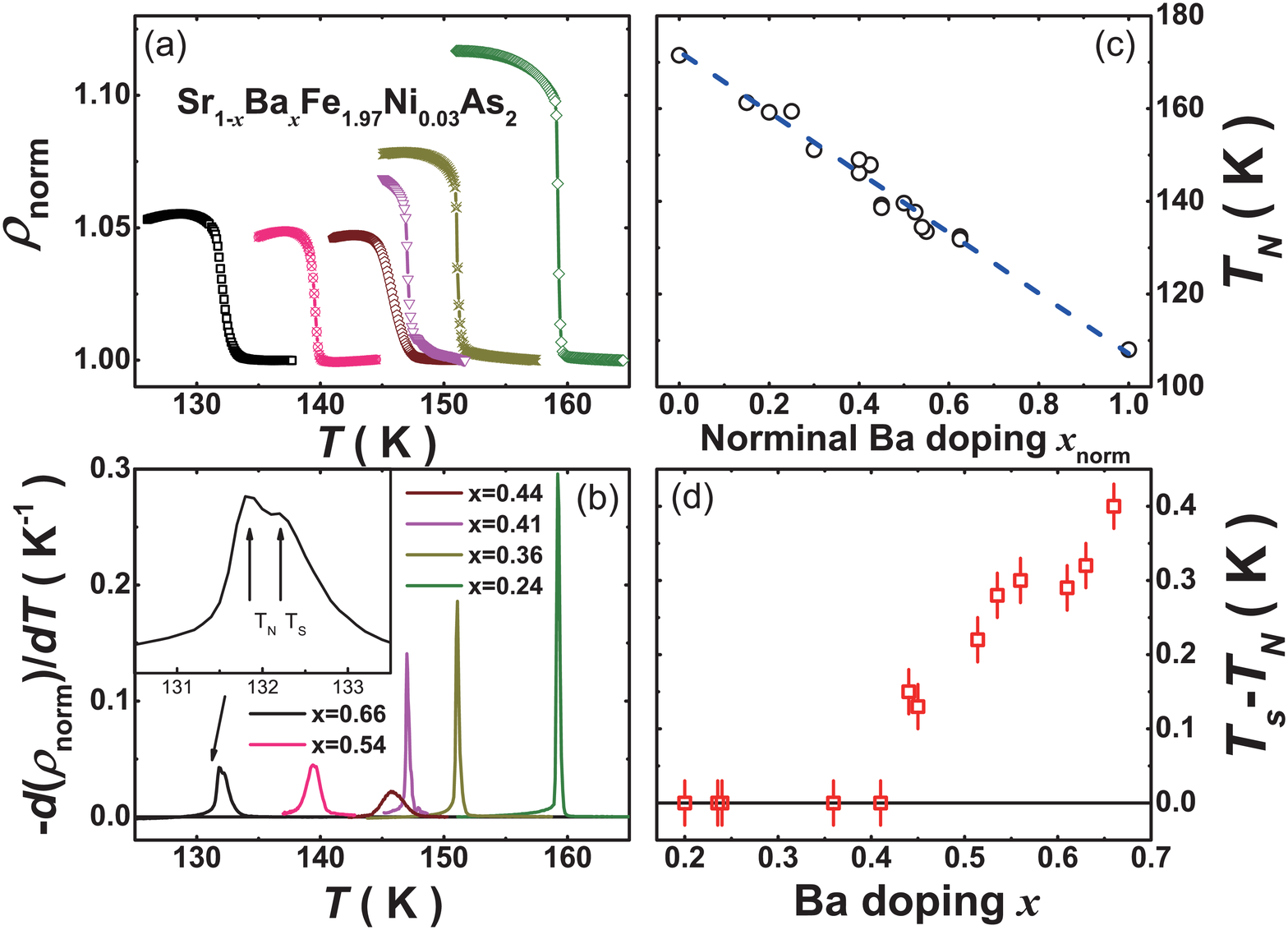}
\caption{(a) Normalized in-plane resistivity $\rho_{\mathrm{norm}}$ of Sr$_{1-x}$Ba$_{x}$Ni$_{0.03}$Fe$_{1.97}$As$_2$. The values are normalized to those at about 5 K higher than $T_N$ for convenience. (b) Temperature dependence of $-d(\rho_{\mathrm{norm}})/dT$ showing sharp and broad peaks around $T_N$ at low and high doping levels, respectively. In sufficiently high doping samples such as $x =$ 0.66, the splitting between the magnetic and structural transitions results in two peaks as shown in the inset. (c) Relationship between $T_N$ and nominal doping $x_{\mathrm{norm}}$. The dashed line is a linear fit to the data. (d) Doping dependence of $T_s-T_N$. The error bars are manually set to 0.03 K, which is much larger than fitted error bars from the two Lorentz fitting as described in the main text.}
\label{fig1}
\end{figure}

\section{RESULTS AND DISCUSSIONS}
Figure 1(a) gives the temperature dependence of the in-plane resistivity $\rho_{\mathrm{norm}}$ normalized by its value at about 5 K above the AF transition temperature $T_N$. A jump around $T_N$ for lower doping samples can be easily seen, suggesting the first order nature of the transitions. As shown previously \cite{PrattDK09,ChenY11}, the precise values of $T_N$ and the structural transition temperature $T_s$ can be determined from the first derivative of $\rho_{\mathrm{norm}}$ [Fig. 1(b)]. For lower doping samples where both transitions are first order and happen at the same temperature, a very sharp peak appears. When the magnetic and structural transitions are well separated, two peaks can be observed corresponding to $T_N$ and $T_s$ as shown in the inset of Fig. 1(b). In the intermediate region where the separation is not obvious, the broad peak can be fitted with two Lorentz functions with the same width, whose peak positions are labeled as $T_N$ and $T_s$. It should be noted that these values show little change if one uses two Gaussian functions to fit the data. Figure 1(c) shows that $T_N$ linearly depends on the nominal Ba doping $x_{\mathrm{norm}}$, but deviation occurs even for samples with the same $x_{\mathrm{norm}}$. This is most likely due to inhomogeneity during the crystal growth process. Therefore, we calculate the Ba doping level $x$ in the following according to its $T_N$. The doping dependence of $T_s-T_N$ is shown in Fig. 1(d), which suggests that the separation between the two transitions becomes non-zero above $x =$ 0.41 and then monotonically increases with increasing $x$.

\begin{figure}
\includegraphics[scale=0.225]{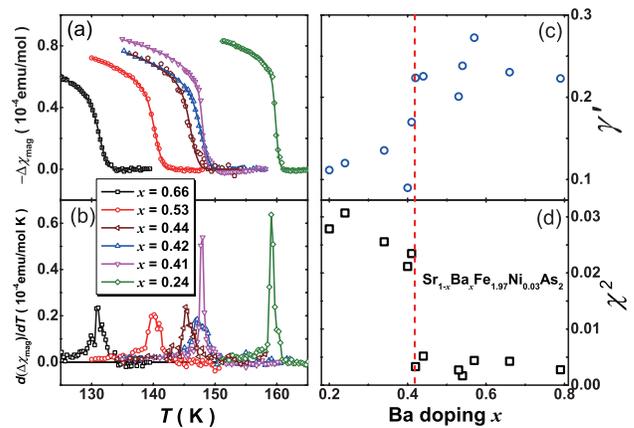}
\caption{(a) Temperature dependence of subtracted magnetic susceptibility $-\Delta \chi_{\mathrm{mag}}$. Here $\Delta \chi_{\mathrm{mag}} = \chi_{\mathrm{mag}}-\chi_{bg}$, where $\chi_{bg}$ is the linear background. The solid lines are fitted as described in the main text. (b) Temperature dependence of $d(\Delta \chi_{\mathrm{mag}})/dT$. (c) and (d) Doping dependence of $\gamma'$ and reduced $\chi^2$ for the fitting in (a). The vertical dashed line represents the crossover from first order to second order for the magnetic transition.}
\label{fig2}
\end{figure}

Figure 2 gives the results of magnetic susceptibility $\chi_{\mathrm{mag}}$ to determine the nature of the magnetic transition in Sr$_{1-x}$Ba$_{x}$Ni$_{0.03}$Fe$_{1.97}$As$_2$. The magnetic susceptibility for all samples studied here shows a linear temperature dependence above $T_N$ \cite{WangXF09}, which is subtracted from $\chi_{\mathrm{mag}}$ to obtain $\Delta\chi_{\mathrm{mag}}$. As shown in Fig. 2(a), while all the $-\Delta\chi_{\mathrm{mag}}$ increase rapidly with decreasing temperature around $T_N$, the upturn looks sharper for lower doping (higher $T_N$) samples, suggesting a crossover from first to second order transition. This can be seen much clearer from $d(\Delta\chi_{\mathrm{mag}})/dT$ as shown in Fig. 2(b), where the nature of the AF transition can be easily judged from both the height and width of the peak around $T_N$. Accordingly, the crossover from the first to second order transitions happens between $x =$ 0.41 and 0.42.

The nature of the magnetic transition can also be studied by fitting $-\Delta\chi_{\mathrm{mag}}$ below $T_N$ with a function of $\chi_0(1-T/T_N)^{\gamma'}$. To account for the tail behavior above $T_N$, a Gaussian distribution of $T_N$ has been included in the fitting \cite{BirgeneauR73}. The Gaussian width $\sigma$ is between 1K and 2 K for high doping samples, which is consistent with previous studies by neutron diffraction for similar systems \cite{ZhangW16,DhitalC14}. The critical exponent $\gamma'$ for most high doping samples is between 0.2 and 0.25. With deceasing Ba doping, $\gamma'$ becomes much smaller for x $<$ 0.4. More clearly, the reduced $\chi^2$ measuring the goodness of fit increases sharply right below $x =$ 0.41 as shown in Fig. 2(d), which accords with the expectation that the above function cannot fit a first order transition well.

\begin{figure}
\includegraphics[scale=0.225]{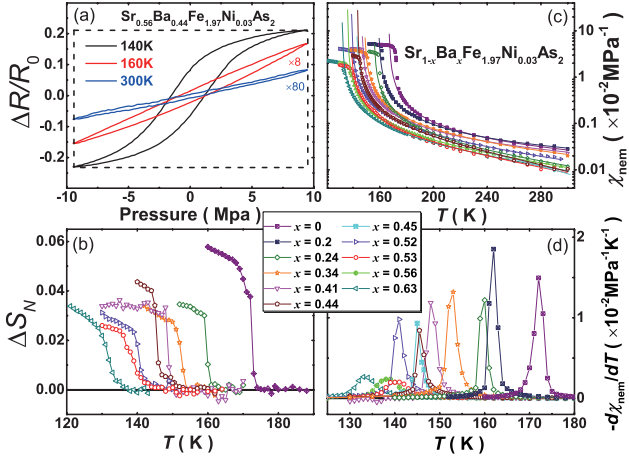}
\caption{(a) Resistivity change $\Delta R/R_0$ under uniaxial pressure along the (110) tetragonal direction for the $x =$ 0.44 sample with $T_N$ = 145.8 K. The dashed rectangular box indicates the total area used to normalized the hysteresis area at 140 K. (b) The temperature dependence of normalized area $\Delta S_N$ with constant high-temperature values subtracted. (c) Temperature dependence of nematic susceptibility $\chi_{\mathrm{nem}}$ with log-scale vertical axis. The solid lines are fitted with the Curie-Weiss-like function as described in the main text. (d) Temperature dependence of $-d\chi_{\mathrm{nem}}/dT$ that shows peak around the transition. }
\label{fig3}
\end{figure}

After determining the nature of the magnetic transitions, we further investigate nematic transitions in these materials by studying nematic susceptibility $\chi_{\mathrm{nem}}$. Here, $\chi_{\mathrm{nem}}$ is defined as $d(\Delta R/R_0)/dp$ as described previously \cite{LiuZ16}, where $R_0$ is the resistance at zero pressure and $\Delta R$ = $R(p)$ - $R_0$. It has been shown that $\chi_{\mathrm{nem}}$ is directly associated with nematic transition in BaFe$_{2-x}$Ni$_x$As$_2$ \cite{LiuZ16}. Figure 3(a) shows some of the raw data for the $x =$ 0.44 sample. At 140 K that is smaller than $T_N$, a ferromagnetic-like hysteresis loop is observed most likely due to the presence of domains as in a ferromagnetic material. We define $S_N$ as the area of the hysteresis loop divided by the minimum rectangular area that contains it as shown in the dashed box in Fig. 3(a). Figure 3(b) shows the temperature dependence of $\Delta S_N$ where the high-temperature constant values due to the intrinsic hysteresis of the piezo-bender has been subtracted \cite{LiuZ16}. A clear jump can be seen for lower-doping samples, suggesting a first order transition.

The temperature dependence of nematic susceptibility $\chi_{\mathrm{nem}}$ is shown in Fig. 3(c). It should be noted that since $\chi_{\mathrm{nem}}$ is not well defined below $T_s$ due to the presence of a ferromagnetic-like hysteresis loop, we force a linear fit to the whole range of the data. Therefore, the data below $T_s$ just represent a trend rather than the actual value of the nematic susceptibility. Similar to $d\chi_{\mathrm{mag}}/dT$, the first derivative of $\chi_{\mathrm{nem}}$ also shows a peak feature around $T_s$ as shown in Fig. 3(d). The difference between sharp and broad peaks clearly suggests the difference between first order and second order transitions.

\begin{figure}
\includegraphics[scale=0.4]{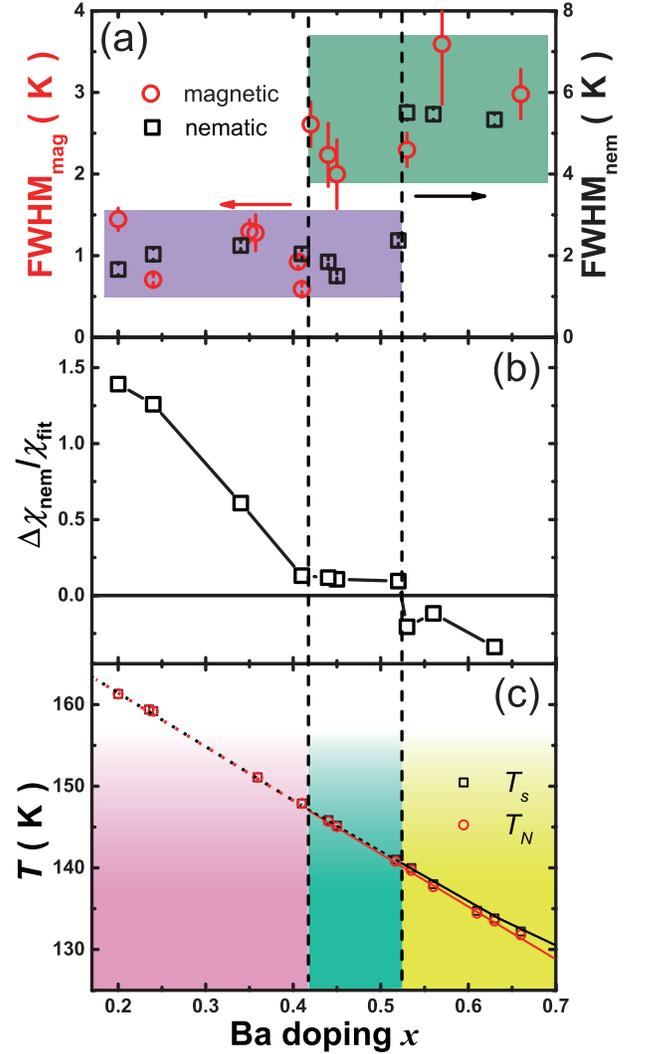}
\caption{(a) Doping dependence of FWHMs of the peaks in $d\Delta \chi_{\mathrm{mag}}/dT$ (red circles) and $-d\chi_{\mathrm{nem}}/dT$ (black squares) for the magnetic and nematic transitions, respectively. The vertical error bars are obtained from the fittings. (b) Doping dependence of nematic susceptibility jump $\Delta \chi_{\mathrm{nem}}/\chi_{\mathrm{fit}}$ at $T_s$. (c) Doping dependence of $T_N$ (red circles) and $T_s$ (black squares). The dotted and solid lines represent first order and second order transitions, respectively. The vertical dashed lines separate the phase diagram into three regions.}
\label{fig4}
\end{figure}

To quantitatively compare the nature of magnetic and nematic transitions, we show the doping dependence of FWHM of the peaks in Figs. 2(b) and 3(d) fitted by a Lorentz lineshape. With increasing Ba doping, the FWHM associated with the magnetic transition jumps to about twice its value above $x =$ 0.41, while a similar jump for the FWHM associated with the nematic transition happens above $x =$ 0.52, as shown in Fig. 4(a). Accordingly, one can identify three regions. In region I (x $\leq$ 0.41), both magnetic and nematic transitions are first order. In region III (x $\geq$ 0.53), both of them are second order. In region II (0.41 $<$ x $<$ 0.53), the magnetic transition becomes second order while the nematic one remains first order, which has never been observed in other materials. Interestingly, FWHM$_{\mathrm{nem}}$ is always about twice as much as FWHM$_{\mathrm{mag}}$ in either region I or region III, which suggests a very close relationship between the AF and nematic orders. 

For a first order nematic transition, one will expect a jump of nematic susceptibility around $T_s$. Such jump is obvious for lower doping samples with strongly first order nematic transition as shown in Fig. 3(c). The amplitude of the jump can be quantitatively analyzed by 
the value of $\Delta \chi_{\mathrm{nem}}/\chi_{\mathrm{fit}}$ at $T_s$, where $\chi_{\mathrm{fit}}$ is the fitted value of a Curie-Weiss-like function as described previously \cite{LiuZ16} and $\Delta \chi_{\mathrm{nem}}$ = $\chi_{\mathrm{nem}} - \chi_{\mathrm{fit}}$. Following the doping dependence of $\Delta \chi_{\mathrm{nem}}/\chi_{\mathrm{fit}}$ at the lower doping regime, it seems that its values will drop to zero at the crossover between region I and II as shown in Fig. 4(b). In region II, the jump becomes very small when the AF transition becomes second order even though the nematic transition itself is still first order. 

Figure 4(c) gives the doping dependence of $T_N$ and $T_s$. The separation between $T_s$ and $T_N$ [see Fig. 1(d)] coincides with the change of the AF transition from first to second order. The change of nematic order from first to second order [Fig. 4(a)] seems to have no effect on the size of the separation between $T_N$ and $T_s$. The magnetic and nematic tricritical points can thus be identified at $x$ of about 0.41 and 0.52, respectively. We note that neutron scattering experiments have reported a few of Kelvins enhancement of $T_N$ and $T_s$ under pressure \cite{DhitalC14,DhitalC12,SongY13,LuX16}, but it will not affect the position of the magnetic tricritical point and the separation of $T_N$ and $T_s$ since they were determined from measurements at zero pressure. Moreover, the pressure has a negligible effect in determining nematic tricritical point here since resistivity measurements under pressure show negligible change of $T_N$ and $T_s$ \cite{FisherIR11} (see also Supplementary Materials \cite{Supplementary}). It is unclear why the results between resistivity and neutron scattering measurements are different, probably because the former use thin slices of crystals that are different from the large and almost square samples used in the latter. 

The rich behaviors of the magnetic and nematic transitions in Sr$_{1-x}$Ba$_{x}$Ni$_{0.03}$Fe$_{1.97}$As$_2$ suggest that the intermediate phase with magnetic and nematic tricritical points is crucial to distinguishing various theories \cite{LvW09,LeeCC09,BrydonPMR11,LeeWC13,KimMG11,FangC08,XuC08,QiY09,FernandesRM11,FernandesRM12,WysockiAL11,KamiyaY11,ApplegateR12}. To our knowledge, our results can only be explained by Ref. \cite{FernandesRM12}. While BaFe$_2$As$_2$ seems to be the case of moderate anisotropy \cite{KimMG11,RotunduCR11}, it has been pointed out that electron-doping results in more anisotropic spin correlations \cite{HarrigerLW09}, suggesting that the phase diagram here is the same as that in Ref. \cite{FernandesRM12} for the strong anisotropy case. Accordingly, the nature of the magnetic and nematic transitions are determined by the nematic coupling, which can be tuned by the change of either Fermi pockets \cite{ZhangY09,YangLX09} or shear modulus \cite{UhoyaWO11,JorgensenJE10}. Moreover, it is predicted that the first order transition of nematic order cannot trigger a first order magnetic transition if the magnitude of the jump of its order parameter at $T_s$ becomes too small, which is also consistent with the results in Fig. 4(b). The excellent consistency between our results and theoretical predictions demonstrates that the nematic order in this system is driven by the spin degree of freedom and suggests the importance of itinerant characteristics of electron system.

\section{CONCLUSIONS}
In conclusion, an intermediate doping region has been unambiguously established in the Sr$_{1-x}$Ba$_{x}$Ni$_{0.03}$Fe$_{1.97}$As$_2$ system by studying the temperature dependence of resistivity and magnetic and nematic susceptibilities. In this region, although the nematic transition is still first order, the jump of the nematic order parameter becomes very small, which coincides with the crossover from first order to second order for the AF transition. Our results agree with the magnetic scenario for an itinerant fermionic model in excellent details. 

\section{ACKNOWLEDGMENTS}
This work is supported by the Ministry of Science and Technology of China (No. 2017YFA0302903, No. 2016YFA0300502, No. 2017YFA0303103, No. 2015CB921302, No. 2015CB921303), the National Natural Science Foundation of China (No. 11674406, No. 11374346, No. 11774401, No. 11374011, No. 11674406, No. 11522435) and the “Strategic Priority Research Program(B)” of the Chinese Academy of Sciences (XDB07020300, XDB07020200, No. 11774401). H. L. and Y. Y. are supported by the Youth Innovation Promotion Association of CAS.

\end{document}